\documentclass[12pt]{article}

\usepackage{amsmath}
\usepackage{amssymb}
\usepackage{graphicx}
\usepackage{subfigure}

\begin{document}

\ \vskip 1.0 in

\begin{center}
 { \large {\bf Quantum Theory, Gravity, and the Standard Model of Particle Physics  }}

\smallskip

{ {\bf  - {\it Using the hints of today to build the final theory of tomorrow} -}}

\vskip 0.2 in

\smallskip

\bigskip

\bigskip

\bigskip

{\large{\bf Tejinder Singh}}

\medskip

{\it Tata Institute of Fundamental Research,}\\
{\it Homi Bhabha Road, Mumbai 400 005, India.}\\
{\tt email: tpsingh@tifr.res.in}\\
\medskip

\vskip 0.5cm
\end{center}

\vskip 1.0 in

\begin{abstract}

\noindent 
When a mountaineer is ascending one of the great peaks of the Himalayas she knows that an entirely new vista awaits her at the top, whose ramifications will be known only after she gets there. Her immediate goal though, is to tackle the obstacles on the way up, and reach the summit. 
In a similar vein, one of the immediate goals of contemporary theoretical physics is to build a quantum, unified description of general relativity
and the standard model of particle physics. Once that peak has been reached, a new (yet unknown) vista will open up. In this essay I
propose a novel approach towards this goal. One must address and resolve a fundamental unsolved problem in the presently known formulation of
quantum theory : the unsatisfactory presence of an external classical time in the formulation. Solving this problem takes us to
the very edge of theoretical physics as we know it today!   

\end{abstract}

\bigskip

\bigskip

\bigskip

\centerline{\it This essay received the Fourth Prize in the Essay Contest `What is ultimately possible in physics?'}
\centerline{\it conducted by the Foundational Questions Institute [http://fqxi.org]}

\newpage


\noindent 

\noindent Modern physics can be said to have begun with the work of Kepler, Galileo and Newton, when the classical laws of motion of bodies were laid down, and the law of gravitation was discovered. The next major development in theoretical physics was Maxwell's theory for the electromagnetic field, and the realization that light is an electromagnetic wave, which travels through vacuum at a universal speed. The inconsistency of this latter result with Newton's mechanics led to the special theory of relativity, and in turn, the incompatibility of special relativity and Newtonian gravitation saw the arrival of the general theory of relativity. Side by side, the failure of classical physics to explain observed phenomena such as the black-body spectrum of electromagnetic radiation, the photo-electric effect, and the spectra of atoms, heralded the discovery of the laws of quantum mechanics. Over the last century or so, quantum theory has been extremely successful in explainning the microscopic structure of matter, and has given us the very precisely tested theories of quantum electrodynamics, electroweak interactions, and strong interactions. To date, the theory has passed every experimental test that has been performed to verify it.

Can we be sure then, that quantum theory is exact, and not an approximation to a deeper underlying theory of mechanics? The answer is no. On the contrary, as will be demonstrated below, one can be sure that the linear quantum theory as we know it, {\it is an approximation} to a nonlinear theory, with the non-linearity becoming significant only near the Planck mass/energy scale! Before we do so, we make two observations. The first is that, contrary to popular perception, quantum theory has not been experimentally tested in all parts of the parameter space quantified by the number of degrees of freedom. The theory is found to work extremely well for atomic systems, say for aggregates having up to a thousand atoms. Also, as we know, it works very well for classical systems, which are aggregates of say $10^{18}$ atoms, or more. In between these two limits, there are some fifteen orders of magnitude (the mesoscopic domain) where quantum theory has not been experimentally tested, simply because the experiments are very difficult to perform. The difficulty lies in isolating the system from the environment; interaction with the environment decoheres the system, and renders it classical. Performing decoherence free tests on mesoscopic quantum systems is a frontline experimental area, and one could be confident that in the next few decades such tests will become possible. As of today, we should be wary of presuming that these experiments will necessarily find that mesoscopic systems do not violate quantum mechanics; in the same spirit that it was wrong to assume that Newtonian mechanics holds for moving objects even if their speed is close to the speed of light. It is entirely possible that mesoscopic objects obey a different mechanics, which reduces to quantum mechanics in the microscopic limit, and to classical mechanics in the macroscopic limit. This may appear counter-intuitive, but it is not something which is ruled out by experiment.

Our second observation has to do with the quantum measurement problem. As is known, during such a measurement, the quantum system makes a transition from being in a superposition of eigenstates of the measured observable, to being in one of the eigenstates, and the probability of any given outcome is proportional to the square of the amplitude for the wave function to be in that state.  If we assume that this transition happens within the framework of standard linear quantum mechanics, then it is explained by the phenomenon of decoherence, in conjunction with the Everett many worlds interpretation. Decoherence destroys interference amongst the various superposed alternatives, while still preserving their superposition. However, since we observe only one of the alternatives in an outcome, we must invoke also the branching of the Universe into many worlds, at the time of a measurement, so that the quantum system, apparatus, and observer, all split into different branches, one branch for every alternative. A difficulty with the many worlds picture is that there is no known way to verify it experimentally, since by definition the different worlds must be non-interacting. If quantum measurement is to be explained without invoking many worlds, the theory must break down during the process of measurement. Such a breakdown is suggested by the reasoning which we now present \cite{singh1}.
  
The evolution of the state of a quantum system with time is of course central to describing the system's dynamics. The concept of time, however, is external to quantum mechanics. It is a classical concept. Time, as a coordinate, is part of a classical spacetime manifold, and on this manifold there resides a classical spacetime metric. The metric is determined by the distribution of classical matter fields, in accordance with the laws of the general theory of relativity. The spacetime manifold and the metric together determine the time evolution of a quantum state. We are thus confronted with a peculiar `fifty-fifty' situation. For its formulation, the quantum theory requires the {\it a priori} presence of classical matter fields; but classical fields are themselves only a limiting case of quantum fields. This is an unsatisfactory, circuitous, state of affairs. A fundamental theory should not have to depend on its own limit, for its formulation. What if there were no classical matter fields in the Universe - a situation entirely possible in principle, and in fact also in practice just after the Big Bang. Something remarkable happens under such circumstances. If there are no classical matter fields, the spacetime metric will undergo quantum fluctuations and will itself not be classical. There is an argument due to Einstein, known as the Einstein hole argument \cite{christian}, according to which, if general covariance is to hold, the point structure of a spacetime manifold is physically meaningful only if the manifold is endowed with a physically determined {\it classical} metric. Now, if the metric is undergoing quantum fluctuations, the underlying classical spacetime manifold is destroyed too. No longer can one talk of evolution with respect to classical time, in quantum theory, if there are no classical matter fields in the Universe. And yet, there ought to exist a formulation of quantum theory - this will be equivalent to the standard formulation, as and when classical matter fields and a classical spacetime exists. We call this the requirement that `there must exist an equivalent formulation of quantum mechanics, which does not refer to an external classical time'. {\it The purpose of this essay is to show that this innocuous requirement has far-reaching consequences, which take us to the physics of limits, and to the limits of physics}!

Consider, as a thought experiment, a Universe which has only quantum mechanical particles, such that their total mass-energy is much less than Planck mass. There is no longer a classical spacetime and no point structure, but in principle there is an equivalent reformulation of the quantum theory. However, the situation is completely different if the total mass-energy becomes comparable to Planck mass. The self-gravity of the system is no longer negligible. As a result the dynamics of the state depends on the state itself, and this makes the dynamics nonlinear. Quantum gravity is decidedly a nonlinear theory. There is now no equivalent reformulation of the linear quantum theory. If an external classical spacetime were to exist, then from the vantage point of that spacetime this nonlinear dynamics will be equivalent to a nonlinear quantum theory, which will reduce to the usual linear theory in the approximation that the total mass-energy of the system is much less than Planck mass. If the total mass-energy far exceeds Planck mass, the nonlinear theory reduces to classical mechanics. The intermediate nonlinear domain in the vicinity of Planck mass is the mesoscopic domain alluded to above. 

Why does this non-linearity not show up when one quantizes a classical theory of gravity, such as general relativity? The answer of course is that linearity is already built into the rules of quantization that are being applied to the classical theory. However, and this is the crucial point, `quantizing' a classical theory of gravity is an illogical step. The rules of quantum theory (such as the Schr\"{o}dinger equation and the accompanying time evolution) are written down assuming an external classical spacetime geometry as being given. If these rules are now applied to quantize the spacetime geometry itself (circuitous reasoning) there is no guarantee that one will arrive at the correct theory of quantum gravity.
Indeed, upon comparing with the previous paragraph we see that the (correctly inferred) feature of non-linearity gets missed out in this traditional approach to quantization.

In order to arrive at a mathematical formulation of quantum mechanics in which there is no external classical time, we start by noting that here one no longer has the point structure of a classical spacetime manifold. Perhaps the most fundamental way of implementing this is to propose that spacetime has become noncommutative, i.e. the coordinates no longer commute with each other. (This is analogous to the situation in ordinary quantum mechanics, where the classical concepts of position and momentum are lost upon imposing the position-momentum commutation relation). We propose that the basic laws of physics are invariant under general coordinate transformations of noncommuting coordinates; thus generalizing the principle of general covariance. In the special case when the total  mass-energy of the system is much less than Planck mass, gravity is
negligible. Here we suggest that the noncommutative spacetime is described by the `noncommutative Minkowski metric' (illustrated here using two
noncommuting coordinates $\hat{x}$ and $\hat{t}$)
\begin{equation}
d\hat{s}^{2} = d\hat{t}^{2} -d\hat{x}^{2} + d\hat{t}d\hat{x} - d\hat{x}d\hat{t}.
\end{equation}            
Suitable noncommutation relations are imposed on the coordinates, and by defining velocity and momenta in a manner analogous to ordinary special relativity, a noncommutative special relativity is constructed. It is shown that this noncommutative special relativity is the new formulation of quantum mechanics which one is looking for, and which becomes the same as ordinary quantum mechanics as and when an external classical spacetime exists \cite{singh1}.

When the total mass-energy of the system becomes comparable to Planck mass, gravity can no longer be ignored, and the above metric must be generalized to a `curved space' metric
\begin{equation}
d\hat{s}^{2} = g_{tt}d\hat{t}^{2} -g_{xx}d\hat{x}^{2} + \theta(d\hat{t}d\hat{x} - d\hat{x}d\hat{t}).
\end{equation} 
The dynamics is now generalized from noncommutative special relativity, to noncommutative general relativity, in a manner analogous to the ordinary commutative case. The spin-off is that one naturally reaches beyond the originally sought for new formulation of quantum mechanics. There results a nonlinear quantum theory, which from the viewpoint of an external classical spacetime is a nonlinear Schr\"{o}dinger equation. The non-linearity has been brought about by the presence of the `curved' metric. In the limit where all metric components approach unity, one recovers linear quantum mechanics. At the other end, as $\theta\rightarrow 0$, spacetime becomes commutative, and one recovers classical mechanics and ordinary general relativity. Since $\theta$ is a function of the ratio $m/m_{Pl}$, where $m_{Pl}$ is the Planck mass, the limit $\theta\rightarrow 0$ corresponds to the total mass energy in the system being much greater than Planck mass. There is a significant departure from linear quantum mechanics and from classical mechanics when $m\sim m_{Pl}$; this is the experimentally untested mesoscopic domain. We are making the case that the transition from quantum to classical mechanics is through a mesoscopic sector which is fundamentally different from either of its limits. This feature has four important consequences, which relate to : (i) the quantum measurement problem, (ii) the origin of black hole entropy, (iii) the cosmological constant puzzle, and (iv) the unification of gravity with the standard model of particle physics.  

\newpage

\noindent {\it The Quantum Measurement Problem} : The requirement that there be a formulation of quantum mechanics which does not refer to an external classical spacetime leads to the conclusion that the theory is an approximation to a nonlinear theory, with the non-linearity becoming important in the vicinity of the Planck mass scale. Let us see what this implies for our understanding of the process of quantum measurement.
Consider a quantum system in a superposition of two eigenstates of an observable. For instance, this could be an electron in a double slit interference experiment, in a superposition of the two states (i) electron passes through the upper slit, and (ii) electron passes through the lower slit. As we know, in the absence of a measuring apparatus (detector) behind the slits, an interference pattern is produced on the photographic plate; this of course is a result of the superposition principle, a hallmark of quantum mechanics. Now consider doing a 
measurement - a (classical) detector is placed behind the upper slit. The relevant state is now the entangled state of the electron and the apparatus. It is the state
\begin{equation}
|\rm{entangled\ state}> = |\rm{electron\ up}>|\rm{detector\ clicks}> \ + \ |\rm{electron\ down}> \ |\rm{detector\ no \ click}>.
\end{equation}   
This is a transitory state, via which the quantum system is going to make a transition to a classical state (electron either goes through
the upper slit or through the lower slit). The entangled state necessarily makes a transition through a mesoscopic phase (microscopic to macroscopic), and hence its evolution is described by a nonlinear Schr\"{o}dinger equation. The entangled state acts as the initial state, which is to be evolved further by the nonlinear equation. The non-linearity breaks the superposition, and only one out of the two outcomes is realized; as
a result the electron goes either through the upper slit, or the lower one, but not both \cite{singh2}.

But we have still not answered a crucial question : what decides which slit the electron will go through?! The answer lies in the nature of the nonlinear terms in the Schr\"{o}dinger equation. These terms contain the phase of the state. It can be shown that depending on the value the phase takes at the onset of measurement, one out of the two superposed states grows exponentially with time, while the other one damps exponentially. Now, since repeated measurements are made at random times, the phase is effectively a random variable. Thus the outcome is random. It can be shown that if an appropriate probability distribution is associated with the random phase, the outcome of the quantum measurement obeys the Born probability rule. That is, the probability of any particular outcome being realized is proportional to the square of the amplitude for the electron to be in the corresponding state. This analysis is easily generalized to the situation when the system is in a superposition of more than two states - depending on the value of the initial phase, one out of the many states grows exponentially, while all others are damped. 

The upshot is that, as a result of the non-linearity, quantum mechanics is a deterministic random theory. The outcome of a measurement is determined by the value of the phase, and the randomness of the phase leads to different outcomes, in consistency with the Born rule. Probabilities are dispensed with, once and for all,  and furthermore, one no longer needs to invoke the untestable `many words interpretation' to `hide' the superpositions which the process of decoherence inherently preserves. Fortunately, the non-linearity is in principle experimentally 
testable - results of quick successive quantum measurements are correlated, unlike in standard quantum mechanics. Also, fundamental constants such as Planck's constant take effective values in the mesoscopic domain which differ significantly from their bare values in the microscopic limit. It is possible that these results can be experimentally tested in the next decade or two. 

\newpage

\noindent{\it The origin of black hole entropy} : 
According to the area theorem in classical general relativity, the sum of the areas of black hole horizons cannot decrease in the interaction of black holes. It is puzzling that a time-reversible mechanical theory such as general relativity should possess such an irreversible feature, which is reminiscent of the second law of thermodynamics. The only plausible answer to this puzzle seems to be that classical general relativity should be thought of as the thermodynamic limit of an underlying quantum statistical theory. This radical rethink is supported by the work of various researchers \cite{various}, and also by the discovery of Bekenstein and Hawking that a black hole possesses an entropy proportional to the area of its horizon.  The underlying statistical theory is assumed to have degrees of freedom, labeled the `atoms of spacetime', which when coarse-grained, give rise to general relativity in the thermodynamic limit. The spacetime metric cannot act as the `atoms of spacetime' [its a thermodynamic variable], in the same way that a thermodynamic variable such as pressure is different from the underlying degrees of freedom such as the momenta of atoms.  

What then is this underlying quantum statistical theory? What are the atoms of spacetime, and what is the origin of black-hole entropy? Our discussion earlier in the article holds the clue.  We have seen that general relativity is the classical limit of a noncommutative gravity theory. These additional noncommuting degrees of freedom (symbolized by the antisymmetric metric component $\theta$)
are the atoms of spacetime. When coarse grained to give rise to a classical spacetime, they also give rise to a thermodynamic interpretation for gravity. The counting of the microstates of the black hole, which account for its thermodynamic entropy, is arrived at via an elegant duality principle, which we now explain.

Consider a particle of mass $m$. According to classical general relativity, its Schwarzschild radius is proportional to its mass. If the particle were to be treated according to the rules of quantum theory, its Compton wavelength is inversely proportional to its mass. The product of the Schwarzschild radius and Compton wavelength is a universal constant, independent of mass, being equal to the square of Planck length. This universal constancy is a puzzle which ought to be explained, since {\it a priori}, general relativity and quantum theory have nothing to do with each other (the former sets $\hbar=0$, while the latter sets $G=0$). The explanation comes from our mesoscopic noncommutative theory, of which both general relativity and quantum theory are limiting cases [the former is the $\theta\rightarrow 0$ limit and the latter is the $\theta\rightarrow 1$ limit]. It is a consequence of the following duality principle which we have proposed and argued for \cite{singh3} :

{\it The strongly gravitational, weakly quantum dynamics of a black hole of mass $M \gg m_{Pl}$ is dual to the strongly quantum, weakly gravitational dynamics of the quantum field theory of a particle of mass $m \ll m_{Pl}$, where $Mm=m_{Pl}^2$}.   

The rationale behind this duality principle is that it maps the Schwarzschild radius of $M$ to the Compton wavelength of $m$, and the Compton wavelength of $M$ to the Schwarzschild radius of $m$. Since these are the only two fundamental length scales in the dynamics, one expects that
a map which interchanges the two lengths makes the dynamics of $M$ dual to the dynamics of $m$, in the sense that the solutions in one case 
can be mapped to the solutions in the other case.

In order to calculate the entropy of the black hole, we make the natural assumption that the mass of the black hole is made up from the mass quanta $m$ of the dual quantum field, there being $N=M/m=M^{2}/m_{Pl}^{2}$ such quanta. The permissible energy levels for the quantum field extend up to Planck mass, in steps of $m$ [as opposed to being a continuum], and those for the black hole extend up to $M$, there hence being $N=M/m=M^{2}/m_{Pl}^{2}$ such eigenstates. The entropy of the black hole is naturally defined as the logarithm of the number of ways in which the $N$ mass quanta can be divided amongst the $N$ eigenstates, and is easily shown to be proportional to the area of the black hole
\cite{singh4}.

The thermodynamic nature of classical gravity can hence to be traced to its underlying noncommutative nature.
It follows from the above analysis that the existence of a black hole entropy, proportional to the area of the black hole horizon,  is a consequence of requiring that there be a new formulation of quantum mechanics which does not refer to an external classical time.  

\bigskip

\bigskip

\noindent{\it The cosmological constant puzzle} : There is definite observational evidence that the Universe is currently undergoing an accelerated phase of expansion. Since normal gravity would cause the expansion to decelerate, the observed acceleration requires us to revise our model of cosmology. Either classical general relativity has to be replaced by a new law of gravity at large scales, or one has to invoke the existence of a new form of matter, which has negative pressure. This new form of matter has come to be called dark energy. The explanation which fits the observational data the best is to have a form of dark energy known as the cosmological constant. This is the so-called $\Lambda$-CDM model. The
cosmological constant is a term in the Einstein equations, proportional to the metric; it is equivalent to a dark energy for which
the negative pressure is exactly equal to the positive energy density, in absolute magnitude.

While the inclusion of the cosmological constant $\Lambda$ is the explanation for the observed acceleration which fits data the best, it gives rise to two very challenging theoretical puzzles. The first is its extreme smallness. The constant has dimensions of inverse length squared, and a measured value of $10^{-56}$ $cm^{-2}$. The natural theoretical length scale for the constant comes from Planck length, so that the constant should have had the value $L_{p}^{-2}$, which is $10^{66}$ $cm^{-2}$. Thus, its observed value, $10^{-122}$, is extremely small,  when expressed in dimensionless natural units. This is the fine tuning problem - why is $\Lambda$ non-zero, and yet so small? The problem is rendered even more severe by the following additional feature : the zero point energy of quantum fields contributes to gravity in the same way as $\Lambda$ does.
Thus the two contributions to the semiclassical Einstein equations, one from the `bare' $\Lambda$, and the other from zero point energy (dressing of the bare $\Lambda$), must finely balance so as to leave just this very tiny residue. Why should this be so?

The second cosmological constant puzzle is known as the cosmic coincidence problem. The observed value of the energy density of $\Lambda$ is of the order of the square of the present value of the Hubble parameter, and comparable to the current matter density. Why should $\Lambda$ be of the order of the matter density {\it today}, where today refers to the epoch when galaxy formation is taking place. $\Lambda$ could as well have been of the order of
the matter density that prevailed much in the past, or that will arise much in the future. In other words, why does $\Lambda$ pick up this particular small value, and not any other?            

Once again, our mesoscopic nonlinear quantum theory comes to the rescue, and explains these two theoretical puzzles \cite{singh5}.
 In the noncommutative gravity theory, the $\Lambda$ term takes the form $\Lambda(g_{ik} + \theta_{ik})$. In the microscopic limit this $\Lambda$ has the interpretation of zero point energy of quantum fields, and in the macroscopic limit it has the interpretation of the bare cosmological constant. There is indeed only one $\Lambda$, and hence the question of finely balancing a bare $\Lambda$ and a dressed $\Lambda$ 
 does not arise. The actual value of the $\Lambda$-term can be deduced from the aforementioned duality principle. In the limit in which the macroscopic black hole could be thought of as the entire observed Universe, which has a mass
$H_{0}^{-1}$ in natural units, the dual quantum field has particles of mass $H_{0}$. The zero point energy of these modes, when added in steps of
$H_{0}$, and extended up to Planck energy, gives rise to an energy density for $\Lambda$ which is order $H_{0}^{2}$, as desired. This solves the fine tuning problem; $\Lambda$ is not a Planck scale quantity, but much larger, simply because we were not counting its contribution to gravity correctly. The cosmic coincidence problem is solved, because our argument holds at any epoch.
$\Lambda$ is no longer a constant; it is an evolving parameter of the order of $H^{2}$ [the Hubble parameter at any given epoch].
This is a prediction which Sorkin \cite{sorkin} has called the ever-present $\Lambda$ and which will be tested by future observations.

\bigskip

\bigskip

\noindent{\it The unification of gravity with the standard model of particle physics} : We now come to the final and perhaps the most fascinating part of this odyssey. This has to do with the unification of gravity with the standard model of particle physics - a topic we have not touched thus far, having restricted ourselves to quantum gravity. The symmetry group of general relativity is of course the diffeomorphism group of the spacetime manifold. The symmetry group of the standard model is the group $U_{1}$ $\times$ $SU_{2}$ $\times$ $SU_{3}$ of local gauge transformations. The symmetry group $G$ of the total Lagrangian of general relativity and the standard model is the semi-direct product of these two groups, just as the Poincare group is the semi-direct product of translations and Lorentz transformations.

It will be great for the sake of unification if there were to exist a space $X$ for which the diffeomorphism group were to be $G$ - then one could say that the unified interaction is geometric; being nothing other than the gravity on $X$. Unfortunately, there are theorems which say that this is impossible, because of the semi-direct product structure of $G$. Enter noncommutative geometry! It turns out that {\it if} $X$ is a noncommutative space, such a diffeomorphism group does exist. $X$ is then the product M $\times$ F of an ordinary manifold $M$ and a finite noncommutative space $F$. A suitable choice of $F$ then represents the standard model of particle physics \cite{chamseddine}; many of whose properties are predicted as a consequence of noncommutative geometry. This of course is very elegant and gratifying - one is closer to a geometric description of unification than ever before. 

The above scenario is however completely classical. It is assumed that a quantum theory of the unified interactions will be arrived at by quantizing the above classical theory in the usual way. This is where we differ from Chamseddine and Connes. Our philosophy in this essay has been that a noncommutative theory of gravity is inherently quantum, once suitable noncommutation relations have been imposed on the coordinates. In the same spirit, the unified theory on the space $X$ above is to be considered as inherently quantum, after imposing noncommutation relations on the coordinates on this extended space. As before, there is no scope for an external classical time in the quantum theory. In fact, as Connes has emphasized \cite{chamseddine}, and this is very remarkable, there is a `God-given' direction of time in noncommutative geometry.
 At the Planck scale, this will be a nonlinear quantum theory. At lower energy scales, the quantum theory becomes linear, and gravity becomes classical. This is consistent with the picture we have developed in this essay.

This is as far as we have come up until now. More remains to be done to make the framework concrete and complete. But the clues are
tantalizing. We introduced noncommutative geometry to address a problem in quantum theory - the problem of removing classical time.
Chamseddine and Connes very elegantly use noncommutative geometry to put forth a unified geometric picture of fundamental interactions.
Putting this all together, we see that the scope of noncommutative geometry is grand - it is the next natural step in the generalization of geometry, beyond the work of Riemann. The new geometry not only seems to provide a platform for unification, but also addresses perplexing
issues in quantum theory. It has the potential to provide us with a quantum unified description of all fundamental interactions, the goal
we set out towards, in this essay.  

\bigskip

\bigskip

\centerline {\it The view from the summit}

\bigskip
 
Only after we have the unified theory, and only after we have worked through it, will we know what more secrets nature holds for us. As of now, we can only speculate : time-machines, baby universes, faster than light travel, closed time like curves, naked singularities, neutralinos, new elementary particles,...? As physicists, we first need a concrete theory in hand, and the technology to go with it,  before we fantasize about `ultimate possibilities'. [The Greeks had great astronomers of their time, who knew of the planet Mercury, but they did not figure that the perihelion of Mercury precesses. They did not have the technology to make such a measurement, and they did not have the theory to provide the correct explanation for it].

Perhaps its fair to say that in the end Einstein will turn out to be the winner after all. Unification will be achieved by a generalization of
general covariance to the noncommutative case - his passion for a geometric unification of interactions will be realized, though perhaps not quite in the way he might have imagined. A deep incompleteness in our understanding of quantum mechanics - something which bothered him all his 
life - will have been removed. Goodbye probabilities; welcome back, determinism. Let us end with the words of Einstein \cite{einstein}: 

\smallskip

{\it "There is no doubt that quantum mechanics has seized hold of a beautiful element of truth and that it will be a touchstone for a future theoretical basis in that it must be deducible as a limiting case from that basis, just as electrostatics is deducible from Maxwell equations of the electromagnetic field or as thermodynamics is deducible from statistical mechanics. I do not believe that quantum mechanics will be the starting point in the search for this basis, just as one cannot arrive at the foundations of mechanics from thermodynamics or statistical mechanics ".}

\rightline{- Einstein (1936)}

\end{document}